\documentclass[a4paper]{article}
\usepackage[utf8]{inputenc}
\usepackage[english]{babel}
\usepackage{graphicx}
\usepackage{hyperref}
\usepackage{csquotes}
\hypersetup{
  setpagesize  = false,
  colorlinks   = true,    
  urlcolor     = blue,    
  linkcolor    = black,   
  citecolor    = black    
}
\usepackage{xcolor}
\usepackage{authblk}
\usepackage{geometry}
\usepackage{setspace}
\usepackage{amsmath}
\usepackage{amssymb}
\usepackage{verbatim} 
\usepackage{ragged2e} 
\usepackage{amsfonts}
\usepackage{cancel}
\numberwithin{equation}{section}
\usepackage{physics}
\usepackage{indentfirst}
\usepackage{xcolor}
\usepackage{natbib}

\usepackage{tensor}

\newcommand{\keywordsenglishname}{Keywords}

\renewenvironment{abstract}{%
        \begin{center}
	\begin{minipage}{14cm}
	{\textbf{\abstractname:}}
}{
        \end{minipage}
	\end{center}
}

\newenvironment{keywordsenglish}{
        \def\abstractname{\emph{\keywordsenglishname}}
	\begin{abstract}
}{
        \end{abstract}
}

\title {\textbf{Axion Electrodynamics in Presence of a Stellar Non-homogeneous Background Magnetic Field}}

\author{Bernard Teles de Menezes \href{https://orcid.org/0000-0003-3157-6005}{\includegraphics[scale=0.04]{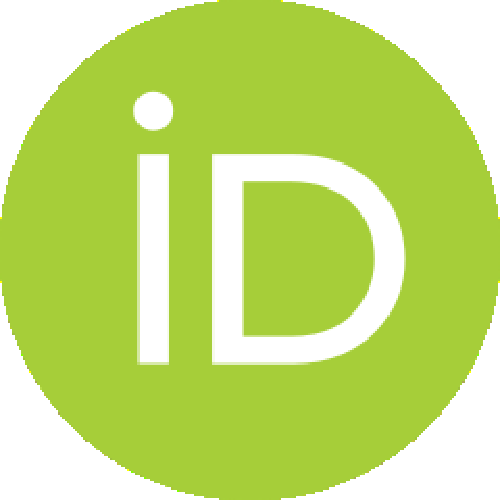}} \thanks{bernardmenezes123@gmail.com} $^1$ }
\affil{$^1$Brazilian Center for Research in Physics, Rio de Janeiro, Brazil.}

\date{\today}

\usepackage{fancyhdr}
\begin{document}

\maketitle
\vspace{6pt}

\begin{abstract}
   We re-assess effects of Axion Electrodynamics in the presence of non-homogeneous background magnetic fields. The Primakoff effect, which is an ALP-photon mixing induced by a background magnetic field, is generalized to the case of non-homogeneous backgrounds. The production of ALPs from a magnetized plasma thermal medium is also computed. 
   The Eddington's solar standard model is adopted as the phenomenological scenario for our settings, and new bounds on the Axion Electrodynamics parameter space are imposed by considering helioseismological and neutrino observational measurements.
\end{abstract}
\vspace{.5cm}
\begin{keywordsenglish}
Axion, Photon, Axion Electrodynamics, Kinetic Mixing, Physics beyond the Standard Model, Dark matter, Strong-CP problem.
\end{keywordsenglish}

\vspace{6pt}



\newpage
\tableofcontents
\vspace{.5cm}

\section{Introduction}

The Standard Model of Elementary Particle Physics (SM) is one of the great successes of modern Physics. Currently, there is a strong consensus within the scientific community that the SM constitutes the fundamental basis of our understanding regarding elementary particles and their interactions. However, the SM cannot answer all questions of fundamental physics, thus requiring new theoretical and experimental perspectives of physics beyond the standard model (BSM). Today, we have major examples of BSM theories, such as Supersymmetry \cite{bilal2001introduction}, Superstrings \cite{schwarz2001introduction}, Loop Quantum Gravity \cite{ashtekar2021short}, etc., which not only inhabit the frontier of elementary particle physics science but also offer new insights and paths toward new physics.

In the formulation of the SM, a fundamental concept is the invariance of physics under a set of transformations called symmetry transformations. Noether's theorem \cite{noether1983invariante} establishes that for every invariance of a system under a continuous symmetry transformation, there emerges a conserved quantity. Thus, the preservation of the physical properties of a given system is associated to a set of specific mathematical operations. A major example is the invariance of Lagrangian densities under charge conjugation (C), parity (P), and time reversal (T) transformations. These transformations are related to the preservation of transition probabilities in certain physical processes. In this context, charge conjugation, which transforms the theory of a physical particle into its antiparticle version, implies that processes involving antiparticles and particles have the same probability. Parity, which inverts spatial directions, implies the spatial isometry of transitions, and time reversal, which inverts the direction of time, results in the invariance of transition probabilities by the exchange of final and initial states. We can also have invariance under consecutive transformations, such as CPT symmetry (the product of C, P, and T transformations in sequence). The latter is closely linked to the preservation of Lorentz symmetry \cite{greaves2014cpt}.

One of the great pillars of the SM is Yang-Mills-Shaw (YM-S) theories \cite{furtado2022yang}, which form the theoretical basis of the fundamental interactions of nature. One of its notorious aspects is the possibility of an interaction term that violates CP symmetry, known in the literature as the $\theta$-term:
\begin{equation}
\label{ThetaTermQCD}
\mathcal{L}_{\theta-term}= \frac{g^2}{32\pi}\theta \ G^{a,\mu\nu}\Tilde{G}^a_{\mu\nu},
\end{equation}
where $g$ is the so-called non-abelian coupling constant, $\theta$ is a dimensionless parameter, and the non-abelian field strength tensor, along with its dual, has the form 
\begin{equation}
\label{NonAbelianFieldStrengthTensor}
    G^a_{\mu\nu} = \partial_\mu A_\nu^a - \partial_\nu A_\mu^a +gf^{abc}A^b_\mu A^c_\nu \ \ \ \ \ \ \text{and} \ \ \ \ \ \ \ \Tilde{G}^a_{\mu\nu} = \frac{1}{2}\varepsilon^{\mu\nu\alpha\beta}G_{\mu\nu}^a.
\end{equation}

In abelian theories, i.e., the Electromagnetic sector, this interaction does not contribute due to abelian Bianchi identities; however, in the case of non-abelian theories, this is no longer true. It is also possible to trace its origin in other BSM theories, such as Supersymmetry \cite{bilal2001introduction}. Historically, CP violation is at the heart of the formation of Electroweak theories \cite{Bjorken:1980gb, QuestionParityLeeYang,glashow1961partial,weinberg1967model,salam1959weak} and currently has extensive experimental verification \cite{ParityMadameWu}. On the other hand, CP symmetry breaking is not observed experimentally in strong interactions \cite{ai2022limits}, and therefore the $\theta$-term should be zero. This is evidenced by the fact that term \eqref{ThetaTermQCD} leads to a non-zero neutron electric dipole moment. This, in turn, depends on the spin vector which is sensitive to CP transformations and can therefore lead to CP symmetry breaking in strong interactions. Experimental tests show that the neutron dipole moment is $d_n < 1.28 \times 10^{-26} e \cdot \text{cm}$, which leads to the parameter $\theta \sim 10^{-10}$. Therefore, if the $\theta$ parameter is non-zero, it needs to be very small. This scenario constitutes what is known in the literature as the \textit{strong CP problem}.

A possible answer to this question was proposed by Peccei and Quinn in 1977 \cite{peccei1977cp}. They proposed promoting the $\theta$ parameter to a new dynamic field, which later came to be called the \textit{axion}. In this way, the energy minimum of Quantum Chromodynamics (QCD), the Yang-Mills theory of the SU(3) color group responsible for describing strong interactions is modified. If we redefine the axion field such that the expected value of its vacuum state is zero, then the term \eqref{ThetaTermQCD} is canceled, thus recovering CP invariance in strong interactions. We thus obtain a new elementary particle, which, due to the nature of the term \eqref{ThetaTermQCD}, is generated by a pseudo-scalar field. 

Motivated by the strong CP problem and by possible applications in the physics of fractional charges with so-called dyons (objects with electric and magnetic charges), Wilczek proposed Axion Electrodynamics \cite{AxionElectrodynamicsWilczek}. Since the $\theta$ parameter was promoted to a dynamic field, the term \eqref{ThetaTermQCD} can now also contribute to the abelian YM-S sector. In the literature, it is common to find the covariant axion-photon interaction term in the form:
\begin{equation}
\label{AxionPhotonCoupling}
    \mathcal{L}_{AX}\bigg |_{\ \text{interaction}} = \frac{g_{a\gamma}}{4}\varphi\Tilde{F}_{\mu\nu}F^{\mu\nu},
\end{equation}
where $\varphi$ is the axion field, and $g_{a\gamma}$ is the axion-photon coupling constant which has the dimension of $\text{length}^{-1}$.  $F_{\mu\nu}$ and $\Tilde{F}_{\mu\nu}$ are the abelian versions of \eqref{NonAbelianFieldStrengthTensor} (in the abelian case, we have only one group index "a" and therefore the group structure constants $f^{abc}=0$). Unlike Maxwell's Electrodynamics, the axionic interaction \eqref{AxionPhotonCoupling} allows the coupling of two photons mediated by a virtual axion. Processes such as direct photon-photon scattering become measurable. 
The coupling \eqref{AxionPhotonCoupling} can be represented by the Feynman diagram:

\begin{figure}[h!]
    \centering
    \includegraphics[scale=.6]{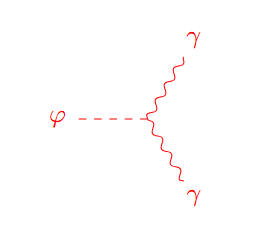}
    \caption{Feynman Diagram of Axion Electrodynamics Interaction.}
    \label{fig:placeholder}
\end{figure}

Thus, the axion proposal ceased to be strictly tied to the strong CP violation issue in strong interactions and began to permeate other areas of physics. The so-called "axion-like particles" (ALPs, to differentiate from axions related to the strong CP problem) have many applications ranging from condensed matter \cite{sekine2021axion} in topological insulators, to cosmological contexts \cite{marsh2016axion}, and in particular, they constitute one of the main candidates for dark matter \cite{jaeckel2010low} in the category of so-called "weakly interacting massive particles" (WIMPs). Many works also explore the cosmological impacts of ALPs, considering large-scale effects in the Universe \cite{berezhiani1991cosmology}\cite{khlopov1998large}.

There are many current experiments seeking to detect axionic particles or ALPs, ranging from microwave cavities in ADMX ("axion dark matter experiment"), searching for influences on astrophysical scales, such as CAST (CERN Axion Solar Telescope) \cite{graham2015experimental}, which seeks to measure axions produced in the solar core through the conversion of X-rays into axions in the presence of external magnetic fields of the order of $9.5 \ T$, or in the context of high-energy particle physics searching for signatures of exotic particles (including dark matter) through the MoEDAL ("Monopole and Exotics Detector at the LHC") \cite{mitsou2021results} and Atlas \cite{atlas2024search} collaborations.

One of the peculiar aspects of Axion Electrodynamics is that we can show that the presence of an external magnetic (and/or electric) field induces axion/photon mixing, i.e., the so-called Primakoff effect \cite{MixingofPhotonwithLowMassParticles}. A similar effect can also be derived from other physical contexts, such as Lorentz symmetry violation \cite{de2026stellar} or a possible mixing with a $U(1)$ dark sector (dark photon) \cite{redondo2008helioscope}. Since ALPs can constitute part of dark matter, the Primakoff effect can be used as a portal to the dark sector, requiring only the presence of a significant external magnetic field. Experiments like CAST use helioscopes to measure the conversion of axions produced in the sun's core into x-rays in the large magnetic fields of the detectors \cite{AxionLikeProduction}. Another very interesting experiment, carried out by the OSQAR collaboration, seeks to measure the transmission of light signals through opaque materials in the presence of magnetic fields, i.e., the so-called "shining light through wall" \cite{ballounew}.

\begin{figure}[h!]
    \centering
    \includegraphics[scale=.23]{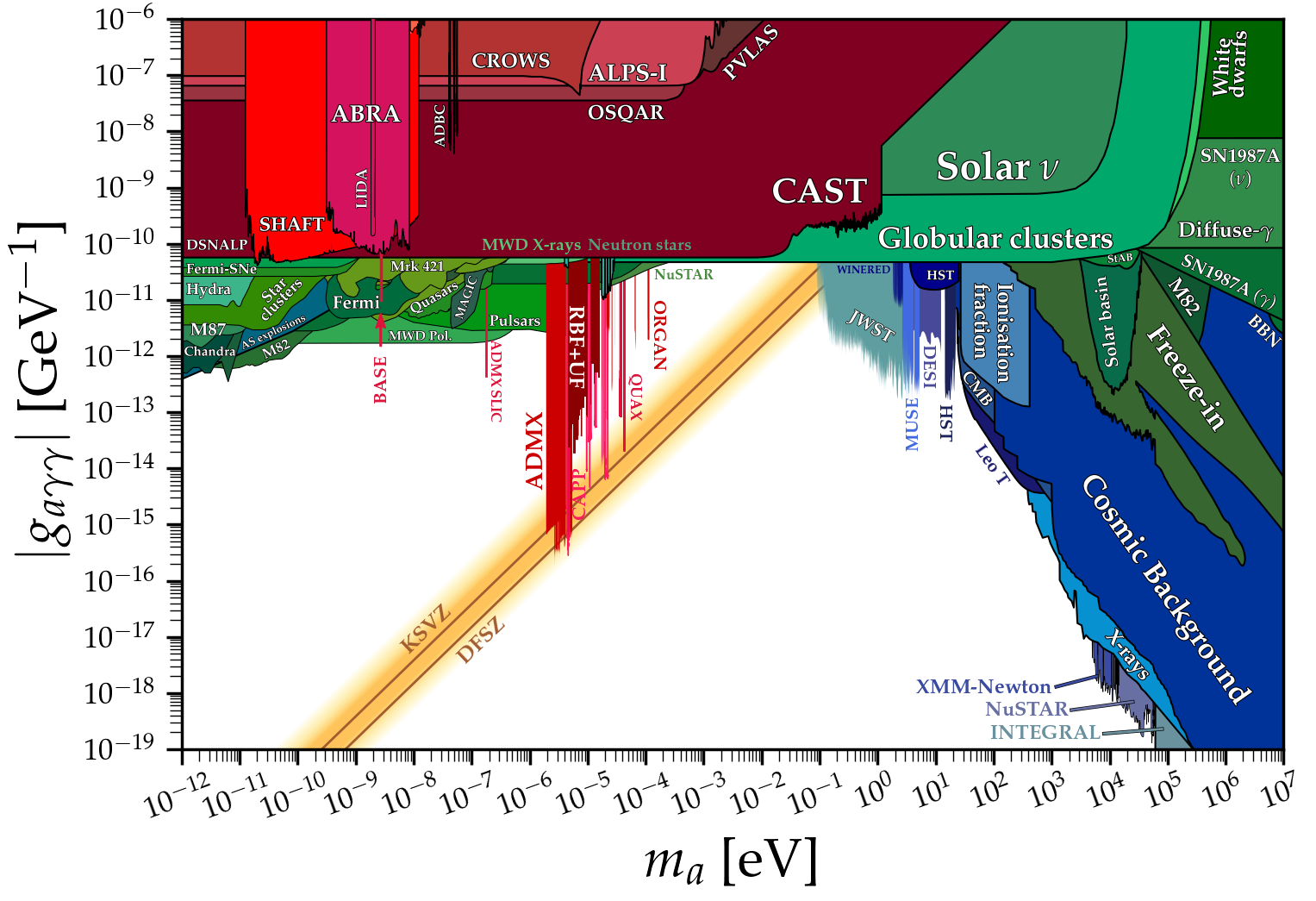}
    \caption{Current parameter space of Axion Electrodynamics. Source: \cite{ciaran2}.}
    \label{EspacoDeParametrosAxion}
\end{figure}

An important point to emphasize is the large possible range for ALP masses. In general, there are two possible perspectives: the first considers scattering processes capable of producing axions at the eV-TeV scale \cite{paixao2024axions}. We have, for example, the reaction $\gamma +\gamma \rightarrow \varphi \rightarrow \gamma +\gamma$ which is capable of producing axions through the Primakoff process, with the axion mass taking values up to the TeV scale \cite{schoeffel2021photon}. For lead-lead collisions, the exclusion limits for the axion mass are in the range $m \simeq (5 - 100) \ GeV$, for a coupling constant $g_{a\gamma} \simeq 0.05 \ TeV^{-1}$, with a confidence level of $95 \ \%$ \cite{d2021collider}. The second considers astrophysical contexts capable of producing axions with an upper bound for their mass in the scale $10^{-10}$ eV - $30$ keV \cite{ayala2014revisiting}. Recent studies for resonant ALP-photon conversion in the magnetar SGR J1745-2900 exclude couplings $g_{ay} > 10^{-12} \ GeV^{-1}$ for $m\leq 10^{-6} \ eV$ \cite{bondarenko2023neutron}. Another well-established limit was obtained by the CAST experiment, which reported a limit of $g_{a\gamma}\simeq 0.66 \times 10^{-10} \ GeV^{-1}$ for $m \leq 0.02 \ eV$, with a 95\% confidence level \cite{cast2017new}. Figure (\ref{EspacoDeParametrosAxion}) schematizes the current state of the art in the investigation of the parameter space of axion electrodynamics, divided among the main experiments and physical contexts where a signature of the axion presence is sought.

This paper is written in the following manner. In section \ref{SunsProfile}, the phenomenological scenario considered in this work is described in details, i.e. the Sun physics and its magnetic field profile. Following section \ref{AxionElectrodynamics}, we demonstrate axion Electrodynamics in the presence of Stellar magnetized plasma, with the photon's dispersion relation in a magnetized plasma and its birefringent effect, i.e. the Faraday Effect. The case of a homogeneous background magnetic field is considered first, and it is followed by the demonstration of the axion-photon kinetic mixing due to the presence of the background magnetic field, i.e. the Primakoff effect. A new method to compute the probability transitions through linearized propagators is also demonstrated, and the traditional method is revised in order to confirm the efficiency of the new method. In section \ref{WKBsolutionInhomogeneousMagneticField}, we develop the same features of axions electrodynamics of section \ref{AxionElectrodynamics}, but now considering the case of a inhomogeneous background magnetic field. As an application, 
we compute the flux of axions produced at the sun, due to the Primakoff effect, and impose bounds on the axion's coupling strength. Since our methods can be applied to an inhomogeneous background magnetic field, the full stellar magnetic field profile can be used. The future perspectives and final remarks on the work are discussed in section \ref{Conclusion}.

\section{Sun's Profile and Magnetic Field}
\label{SunsProfile}

We start by describing the phenomenological scenario considered in this paper. For the developments that follows, we will consider the system of natural units, i.e., $c=(\varepsilon_0\mu_0)^{-1/2}=\hbar=1$, with $\varepsilon_0=\mu_0=1$. The sun physics is modeled by a $n=3$ polytropic model \cite{kippenhahn1990stellar}. This model can be derived from the set of equations:
\begin{equation}
\label{StructureEquations}
    \begin{split}
        &\dv{P}{r}(r)=-\frac{Gm(r)\rho(r)}{r^2}, \\
        & \dv{m(r)}{r} = 4\pi r^2 \rho(r), \\
        & T^3(r) = \frac{(1-\beta)}{\beta}\frac{45}{\mu_n m_p \pi^2 }\rho(r).
    \end{split}
\end{equation}
The first equation of \eqref{StructureEquations} is the spherically symmetric hydrostatic equilibrium condition in the presence of Newtonian gravitational potential (for the Sun, we have the metric deviation $2GM_\odot/R_{\odot}c^2<<1$). The second equation is the spherically symmetric mass continuity equation. The third equation is derived from the Sun's equation of state, assuming the stellar material composed of a gas of ions and photons. The parameter $\beta = P_{gas}/(P_{gas}+P_{rad})$, with $P_{gas}$ the gas pressure and $P_{rad}$ the radiation pressure, is the so called fractional pressure.  

 Assuming the fractional pressure $\beta \approx \textrm{constant}$ through the star, we can propose the polytropic ansatz $P(r) \propto \rho^{1+1/n}$, and with the help of the first two equations of \eqref{StructureEquations}, we can derive the so-called Lane-Emden equation:
\begin{equation}
\label{LaneEdenEq}
    \frac{1}{\xi^2}\frac{d}{d\xi}\left(\xi^2 \dv{\theta(\xi)}{\xi}\right)=-\theta^n (\xi),
\end{equation}
where we used the parametrization $\xi = ar$ and $\rho(\xi)=\rho_0 \theta^n(\xi)$. Using the Sun's equation of state, we can show that $n=3$.

\begin{figure}[h!]
    \centering
    \includegraphics[scale=.8]{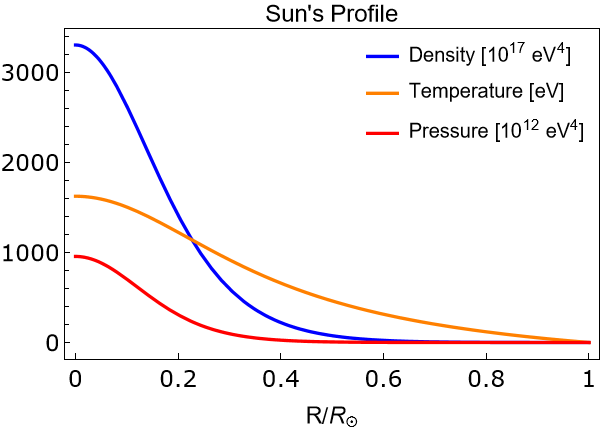}
    \caption{Plot from the Sun's $n=3$ polytropic model profile \eqref{SunProfile} (Author's own production).}
    \label{PlotSunProfile}
\end{figure}

Solving \eqref{LaneEdenEq} (for $n=3$) numerically, gives us the following profile for the Sun (Figure \ref{PlotSunProfile} contains the plot of our derived solution):

\begin{equation}
\label{SunProfile}
    \begin{split}
        \rho(r) &= \rho_0 \  \theta^3(r.a), \\ 
        T(r) &= T_0 \ \theta(r.a), \\ 
        P(r) &= P_0 \theta^4(r.a),
    \end{split}
\end{equation}
with the central quantities (in natural units $\hbar=c=k_b=1$) given by,
\begin{equation}
    \begin{split}
        \rho_0  &= 3300.85 \times 10^{17} \ eV^4 , \\
        T_0 &= 1632.49 \ eV , \\
        P_0 &= 954.88 \times 10^{12} \ eV^{4}.
    \end{split}
\end{equation}

We consider the Sun's radius $R_\odot=6.96 \times 10^8 \ m$ and mass $M_\odot=1.99 \times 10^{30} \ kg$ as input data for the model and we defined $a=b/R_\odot$, such that $b=6.897$ and $\theta(b)=0$. For the sun chemical composition, we considered the abundances in \cite{asplund2009chemical},  which gives us the mean molecular weight $\mu_n\approx 0.6$ and we consider the proton mass $m_p=938.272 \ MeV$. The radial dependency of the density is also transmitted to the electron's number density, i.e. $n_e=n_e(r)$, which also implies a radial dependency on the plasma frequency $\omega_p^2=\omega^2_p(r)$ \eqref{PlasmaFrequency}.

The Sun's magnetic field is prominent in 3 different regions, the radiative zone ($r<0.7 R_{\odot}$), the intermediate tachocline zone ($r\approx 0.7 R_\odot$) and the more exterior convective zone ($r> 0.9 R_\odot$). Following \cite{couvidat2003solar,gough1990effect}, we model the Sun's magnetic field profile as a toroidal field:
\begin{equation}
    B_\phi=a(r)\dv{P_k(\cos{\theta})}{\theta},
\end{equation}
with the spherical coordinates $(r,\theta,\phi)$, $P_k(\cos{\theta})$ the Legendre polynomial of degree k, and $a(r)$ the radial profile of the field. For our intentions and purposes, we consider a $k=1$ dipole field. For the radiative region (for $r\leq r_0= 0.712 \ R_\odot$), we use the following radial model:
\begin{equation}
\label{MagneticFieldProfile1}
    a(r) = (1+\lambda)(1+1/\lambda)^\lambda \left(\frac{r}{r_0}\right)^2\left(1-\left(\frac{r}{r_0}\right)^2\right)^\lambda B_{\textrm{rad}},
\end{equation}
with $\lambda= 1 + 10 r_0/R_\odot$ and $1.95 \times 10^5 \ eV^2 \ \lesssim B_{rad}\lesssim 5.85 \times 10^5 \ eV^2$. 

\begin{figure}[h!]
\label{SunsMagneticField}
    \centering
    \includegraphics[scale=.8]{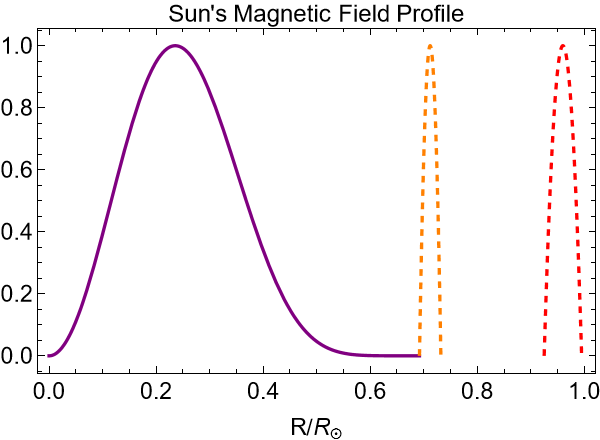}
    \caption{The normalized Sun's magnetic field profile in different regions. The solid purple curve represents the radiative region \eqref{MagneticFieldProfile1}, the orange dashed is the tachocline region and the red dashed is the convective region \eqref{MagneticFieldProfile2} (Author's own production).}
    \label{fig:placeholder}
\end{figure}

For the intermediate tachocline and convective regions, the field profile is given by
\begin{equation}
\label{MagneticFieldProfile2}
    a(r) = \begin{cases}
        &\left[1-\left(\frac{r-r_0}{d}\right)^2\right]B_0, \ \  \textrm{for} \ |r-r_0|\leq d, \\
        & 0,  \qquad \qquad \qquad   \textrm{otherwise},
    \end{cases}
\end{equation}
with $5.85 \times 10^{3} \ eV^2 \lesssim B_0=B_{\textrm{tach}} \lesssim 9.75 \times 10^{3} \ eV^2$ and the half-width $d=0.02 \ R_\odot$ for the tachocline zone. The convective zone is modeled with $0.39 \times 10^3 \ eV^2 \lesssim B_0=B_{\textrm{conv}} \lesssim 0.585 \times 10^3 \ eV^2$ and $d=0.035 \ R_\odot$ for the half-width. The field profile for the 3 regions, normalized by unit, is plotted in Figure \ref{SunsMagneticField}. In next section, we describe the axion electrodynamics in the context of plasma physics.

\newpage
\section{Axion Electrodynamics in the Presence of a Stellar Magnetized Plasma Medium}
\label{AxionElectrodynamics}

In order to compute the production of solar axions in the presence of stellar inhomogeneous fields \eqref{MagneticFieldProfile1}, we start reviewing the case of Axion Electrodynamics in the presence constant homogeneous background magnetic field.  The Axion Electrodynamics Lagrangian \cite{AxionElectrodynamicsWilczek}, in a plasma medium, has the covariant form:
\begin{equation}
\label{AxionEDLagrangian}
    \mathcal{L}_{AX} = -\frac{1}{4}F_{\mu\nu}F^{\mu\nu}- J_\mu A^\mu + \frac{g}{4}\varphi\Tilde{F}_{\mu\nu}F^{\mu\nu} + \frac{1}{2}\partial_\mu \varphi \partial^\mu \varphi - \frac{m^2}{2}\varphi^2,
\end{equation}
where $\varphi$ is the axion field, and the electromagnetic field strength tensor, with its dual, are defined as:
\begin{equation}
\label{EMTensor}
    \begin{split}
        &F^{\mu\nu}=\partial^\mu A^\nu - \partial^\nu A^\mu, \\
        &\Tilde{F}^{\mu\nu}=\frac{1}{2}\varepsilon^{\mu\nu\alpha\beta}F_{\alpha\beta} = \varepsilon^{\mu\nu\alpha\beta}\partial_\alpha A_\beta,
    \end{split}
\end{equation}
In a plasma medium \cite{jackson2021classical} inertial reference frame, we consider a current that obeys Ohm's law:
\begin{equation}
    J^\mu= (0, \sigma\va{E}),
\end{equation}
where $\sigma$ is the electrical conductivity of the plasma in the presence of an electromagnetic wave. 

In a magnetized plasma medium, an electromagnetic wave possesses, in addition to its modified transverse modes, longitudinal modes due to the presence of plasmons, i.e., quasi-quantizations in a magnetized plasma medium. In this work, we will consider only the transverse modes. The reader may check \cite{AxionLikeProduction,terccas2018axion} for future reference.

Consider an electronic plasma in the presence of an external magnetic field $\va{B}_0$. For small electronic motion amplitudes due to the presence of an electromagnetic wave and neglecting collision effects, we can approximate the equations of motion as:
\begin{equation}
    \label{PlasmaAmplitudeEq}
    m_e \partial_t^2\va{x} -e\va{B}_0 \times \partial_t \va{x} = -e\va{E}e^{-i\omega t},
\end{equation}
where we neglect the effects of the wave's magnetic field due to the high intensity of the external field $\va{B}_0$. Considering the temporal dependence of the amplitude in the form $\va{x}(t)=\va{x}e^{-i\omega t}$, expression \eqref{PlasmaAmplitudeEq} can be written as:
\begin{equation}
    \label{PlasmaAmplitudeEqFourier}
    -m_e\omega^2 \va{x} +i\omega e\va{B}_0 \times \va{x} = -e\va{E}.  
\end{equation}
Note that the temporal dependence of $\va{E}$ was also eliminated because it has the same nature as the temporal dependence of the amplitude $\va{x}$. Finding the Green's function associated with \eqref{PlasmaAmplitudeEqFourier}, we can find the electronic amplitude:
\begin{equation}
    \label{ElectronicAmplitudeSolution}
    x^i(t) = \frac{1}{\omega^2-\omega_B^2}\left(\frac{e}{m_e}\delta^{ij}-\frac{ie \omega_B}{\omega}\epsilon^{jik}b^j-\frac{e\omega_B^2}{\omega^2 m_e} b^ib^k\right) E^k e^{-i\omega t},
\end{equation}
such that $b^i=B_0^i/|\va{B}_0|$ and $\omega_B=B_0e/m$ is the so-called cyclotron frequency. Thus, considering the electronic density $\rho_e=-n_e e$, the 3-current density is given by:
\begin{equation}
\label{ConductingCurrents}
    j^i = -n_ee\dv{x^i}{t} = \frac{i\omega_p^2}{\omega^2-\omega_B^2}\left(\omega \delta^{ik} + i\omega_B\epsilon^{jik}b^j-\frac{\omega_B^2}{\omega}b^ib^k\right)E^k e^{-i\omega t}.
\end{equation}
The plasma frequency is defined as
\begin{equation}
\label{PlasmaFrequency}
    \omega_p^2= \frac{n_ee^2}{m_e} = \frac{e^2}{m_e m_p \mu_e}\rho_e,
\end{equation}
with the electron's mass set to $m_e=0.511 \ MeV$, and the electron's mean molecular weight given by $\mu_e=0.18$. 

\begin{figure}[h!]
    \centering
    \includegraphics[scale=.55]{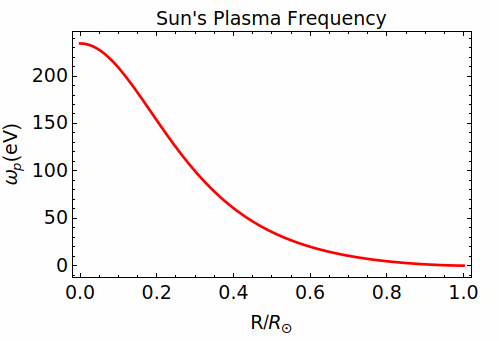}
    \caption{Radial profile of the Sun's plasma frequency, based on the Sun's density profile \eqref{SunProfile}. The plasma is considered to be full ionized (Author's own production}).
    \label{SunsPlasmaFrequency}
\end{figure}

In the context of the Sun, we consider it to be a fully ionized plasma, i.e. $\rho_e=\rho$. Actually, this is an approximation, and for a full treatment we should consider the Saha equation \cite{iben2012stellar}. Since the Sun's Magnetic field is stronger in the radiative zone, and the full-ionized approximation fails at the Sun's external layers, we consider the effects from the Saha equation as small corrections to our results.  The radial profile of the Sun's plasma frequency is plotted in figure \ref{SunsPlasmaFrequency}.

For a plasma, the contribution for the eletric permeability $\varepsilon^{ik}$ comes from the conducting currents \eqref{ConductingCurrents} \cite{liu1973theory}. From the Bohr–Van Leeuwen theorem, the thermal average of the classical magnetization is equal to zero \cite{van1921problemes}. The contribution to the magnetization comes solemnly from the quantum mechanical spin. The thermal average of the spin magnetic moment \cite{kardar2007statistical}: 

\begin{equation}
    \expval{\mu_s} = \frac{\int^1_{-1}d(\cos{\theta}) \mu_s \cos{\theta e^{\mu_s B\cos{\theta}/k_bT}}}{\int^1_{-1}d(\cos{\theta})  e^{\mu_s B\cos{\theta}/k_bT}} = \mu_s \left(\coth{\frac{\mu_s B}{k_bT}}-\frac{k_bT}{\mu_s B}\right).
\end{equation}
Considering the physics of plasmas, we have $k_bT >> \mu_s B$, which gives us the magnetization,
\begin{equation}
    M = n\expval{\mu_s}\approx n \frac{\mu_s^2 B}{3k_b T},
\end{equation}
and consequently the magnetic permeability,
\begin{equation}
    \mu = 1 + \pdv{M}{B}\bigg|_{B=0} \approx 1+\frac{n\mu_s^2}{3k_b T}.
\end{equation}
We conclude that the contribution to the magnetic permeability from the plasma is very small, and for all intentions and purposes here we can assume $\mu\approx1$.

By Ohm's law of conductivity, we see that the electrical conductivity, and consequently the refractive index of the plasma, depends on the direction of the external magnetic field, thus evidencing that the external field induces birefringence in the plasma. The refractive index has the form:
\begin{equation}
    \boxed{\varepsilon^{ik}=(n^2)^{ik}= \delta^{ik} -\frac{\omega_p^2}{\omega^2(\omega^2-\omega_B^2)}(\omega^2\delta^{ik}+i\omega_B \omega \varepsilon^{jik}b^j -\omega_B^2b^ib^k).}
\end{equation}

For a homogeneous external magnetic field, the wave equation, in Fourier space, can be written as
\begin{equation}
\label{WaveEqInPlasmaMagneticField}
	[(\omega^2-|\vec{k}|^2)\delta^{ik}-\Delta_B^{ik}]E^k=0,
\end{equation}
where $\Delta_B$ are the birefringence effects induced by the external magnetic field in the presence of the plasma medium. The spatial indices i, k indicate the action of birefringence on different components of the electromagnetic field polarization vector $\va{E}$. In the context of high frequencies, i.e., $\omega >> \omega_B$, we can write the wave equation for the different polarization modes, in the plasma medium, as:
\begin{equation}
    \label{modifiedDispersionRelations}
     [(\omega^2-\omega_p^2- |\va{k}|^2)\delta^{ik} - \Delta_B^{ik}]E^k =0,
\end{equation}
where, up to first order in $\omega_B/\omega$, the birefringence effects assume the form:
\begin{equation}
\label{cyclotonFrequencyDispersion}
    \Delta_B^{ik} \cong i\omega_p^2 \frac{\omega_B}{\omega}\epsilon^{jik}b^j.
\end{equation}
Considering the Sun's magnetic field profile \eqref{SunsMagneticField}, the cyclotron frequency in the radiative zone ranges from $0.12 \ eV \lesssim \omega_B \lesssim 0.35 \ eV$. In the tacholine region, it ranges from $ 0.35 \times 10^{-2} \ eV \lesssim \omega_B \lesssim 0.6 \times 10^{-2} \ eV$, and in the convective zone it ranges from $0.23 \times 10^{-3} \ eV \lesssim \omega_B \lesssim 0.35 \times 10^{-3} \ eV$.

The changes in the dispersion relation present in \eqref{modifiedDispersionRelations} are not diagonal in the polarization vector space, indicating that the different modes propagates with different dispersion relations. In the next section, we discuss the physics behind this feature. 

\subsection{Faraday Effect in the Presence of Homogeneous Magnetic Field}

Another effect present in a plasma medium with an external magnetic field is the rotation of the polarization angle as the wave propagates in the plasma, i.e., the so-called Faraday rotation {\color{red}\cite{liu1973theory}}. For this, let us consider the solution \eqref{ElectronicAmplitudeSolution} in a situation where the external magnetic field is parallel to the propagation direction of the wave, i.e., $\va{B}_0 \ || \ \va{k}$. It is convenient to assume the conditions:
\begin{equation}
    \label{PolarizationSolution+-}
    \begin{split}
        \va{E}_\pm(t) &= E_\mp \va{\epsilon}_\pm e^{-i\omega t}=\frac{(E_1 \mp iE_2)}{\sqrt{2}} \va{\epsilon}_\pm e^{-i\omega t}, \\
        \va{x}_\pm(t) &= x_\mp \va{\epsilon}_\pm e^{-i\omega t}= \frac{(x_1 \mp ix_2)}{\sqrt{2}} \va{\epsilon}_\pm e^{-i\omega t},
    \end{split}
\end{equation}
with the left/right circular polarization vectors,
\begin{equation}
    \va{\epsilon}_\pm = \frac{\va{\epsilon}_1 \pm i \va{\epsilon}_2}{\sqrt{2}}, \ \ \ \ \ (\va{\epsilon}_\pm)^* \cdot \va{\epsilon}_\mp = \va{\epsilon}_\pm  \cdot \va{k}=0, \ \ \ \ (\va{\epsilon}_\pm)^* \cdot \va{\epsilon}_\pm = 1.
\end{equation}
Substituting \eqref{PolarizationSolution+-} into \eqref{PlasmaAmplitudeEqFourier}, we obtain as a solution the electronic amplitude:
\begin{equation}
    \va{x}_{\pm}(t) = \frac{e}{m\omega(\omega \pm \omega_B)}\va{E}_\pm,
\end{equation}
and consequently the electrical permittivity,
\begin{equation}
\label{DispersionRelationCircularlyPolarization}
    \varepsilon_{\pm} = 1 - \frac{\omega_p^2}{\omega(\omega \mp \omega_B)}.
\end{equation}
Therefore, the wave dispersion in the plasma, where $\va{B}_0 \ || \ \va{k}$, is given by:
\begin{equation}
    \omega^2 \varepsilon_{\pm} - |\va{k}|^2 = \omega^2\left(1 - \frac{\omega_p^2/\omega}{(\omega\pm\omega_B)}\right) - |\va{k}|^2 = 0.
\end{equation}
In the high-frequency limit, i.e., $\omega >> \omega_p$, we can write the wave vector for the two polarizations:
\begin{equation}
\label{PlasmaDispersion}
    |\va{k}_\pm| = \omega \sqrt{\left(1 - \frac{\omega_p^2/\omega}{(\omega\pm\omega_B)}\right)} \simeq \omega \left(1 - \frac{\omega^2_p}{2\omega^2}\left[1\mp \frac{\omega_B}{\omega}\right]\right) = \omega \ -\frac{\omega_p^2}{2\omega}\pm \frac{\omega_p^2}{2\omega^2}\omega_B= |\va{k}_0|\pm \Delta k. 
\end{equation}
Thus, the left/right circular polarizations of an electromagnetic wave propagating in a magnetized plasma have different wave vectors. Since a linearly polarized wave can be written as a superposition of circularly polarized waves:
\begin{equation}
    \begin{split}
    \va{E} &= E_0 (\va{\epsilon}_+e^{-i(\omega t - \va{k}_+ \cdot \va{x})} + \va{\epsilon}_-e^{-i(\omega t - \va{k}_-\cdot \va{x})}). \\
    \end{split}
\end{equation}
Taking the real part of the wave:
\begin{equation}
    \Re{\va{E}} = 2E_0 \cos{(\omega t - \va{k}\cdot \va{x})}\bigg[ \cos{(\Delta \va{k}\cdot \va{x})} \va{\epsilon}_1 +  \sin{(\Delta \va{k}\cdot \va{x})}\va{\epsilon}_2\bigg].
\end{equation}
Therefore, the polarization angle of the wave is given by:
\begin{equation}
    \varphi = \tan^{-1}{\left(\frac{E_2}{E_1}\right)} = \Delta k|\va{x}|. 
\end{equation}
We see, therefore, that the polarization angle rotates as the wave propagates in the magnetized plasma. This effect is known in the literature as Faraday rotation. A linearly polarized wave has its polarization vector rotated when traversing the magnetized thermal medium. 

\subsection{Background Field Approximation and the Axion-Photon Mixing}
\label{PrimakoffAxionED}

A very common approach is to expand the fields of the theory as perturbations around a fixed background field. In the context of stellar plasma, much of the contribution to the Primakoff effect comes from magnetic fields produced by thermal photons, electric fields produced by charged plasma particles, and axion production from the strong interaction via nuclear processes \cite{andriamonje2007improved}. However, we can show that the contribution of the first case is dominant. 

From here on, we will consider fields obeying the modified dispersion relations \eqref{modifiedDispersionRelations} due to the source terms of the magnetized plasma medium. We are interested in expanding around a constant background magnetic field:
\begin{equation}
    F^{\mu\nu}= f^{\mu\nu} + F_B^{\mu\nu}, \ \ \ \Tilde{F}^{\mu\nu} = \Tilde{f}^{\mu\nu} + \Tilde{F}_B^{\mu\nu}.
\end{equation}
The background field $F_B^{\mu\nu}$ has only one non-zero combination:
\begin{equation}
\label{BackgroundApproximation}
    F_B^{ij} = -\varepsilon_{ijk}B_b^k  \ \ \ \ \text{and} \ \ \ \ \Tilde{F}_B^{i0}=-B^i_b.
\end{equation}
In this approximation, we neglect terms higher than quadratic in the fields. Substituting \eqref{BackgroundApproximation} into \eqref{EMTensor}, we obtain the Lagrangian:
\begin{equation}
\label{BackgroundApproximationLagrangian}
    \mathcal{L} = -\frac{1}{4}f_{\mu\nu}f^{\mu\nu}-J_\mu A^\mu +\frac{g}{2}\varphi f_{\mu\nu}\Tilde{F}_B^{\mu\nu} + \frac{1}{2}\partial_\mu \varphi \partial^\mu \varphi - \frac{m^2}{2}\varphi^2,
\end{equation}
where we neglect first-order terms in the fields as they do not contribute to the field equations.
The field equations become:
\begin{equation}
\label{EquationsWithBackground}
    \begin{split}
        &\partial_\mu f^{\mu\nu} = g\partial_\mu \varphi \Tilde{F}^{\mu\nu}_B + J^\nu, \\
        & \partial_\mu \Tilde{f}^{\mu\nu}=0, \\
        & (\Box +m^2)\varphi = \frac{g}{2}\Tilde{F}^{\mu\nu}_Bf_{\mu\nu}.
    \end{split}
\end{equation}

We will now show that the background magnetic field indeed induces a mixing between the axion and the photon \cite{MixingofPhotonwithLowMassParticles}. In terms of the perturbative photon field $a^\mu=(a^0, \Vec{a})$ and the background magnetic field $\Vec{B}_b$, the field equations can be written as:
\begin{equation}
    \begin{split}
        &\Box a^\nu = g \partial_\mu \varphi \Tilde{F}_B^{\mu\nu}+ J^\nu \\
        & (\Box +m^2)\varphi= g\Tilde{F}_B^{\mu\nu} \partial_\mu a_\nu,
    \end{split}
\end{equation}
where we assume the Lorenz gauge condition, i.e., $\partial_\mu a^\mu=0$. In terms of the potentials:
\begin{equation}
\label{LorenzGaugeFieldEquations}
    \begin{split}
        &\Box a^0 = -g \vec{\nabla}\varphi \cdot \Vec{B}_b, \\
        &\Box \Vec{a} = g \partial_t \varphi \Vec{B}_b+\vec{J},\\
        &(\Box + m^2)\varphi =-g \partial_t \Vec{a} \cdot \Vec{B}_b - g \vec{\nabla}\varphi \cdot \Vec{B}_b. 
    \end{split}
\end{equation}
Now, to expose the kinetic mixing, we need to consider the field equations in 4-momentum space $k^\mu=(\omega,\Vec{k})$. Photons obey the dispersion relations \eqref{modifiedDispersionRelations}, and axions:
\begin{equation}
    k_\mu k^\mu=\omega^2-|\Vec{k}|^2=m^2.
\end{equation}
Assuming the Coulomb gauge ($\div{\vec{a}}=0$), and a homogeneous external magnetic field,  we can rewrite equations \eqref{LorenzGaugeFieldEquations}, in 4-momentum space, as:
\begin{equation}
\label{FourierFieldEquations}
    \begin{split}
        & ((\omega^2 -\omega^2_p -|\Vec{k}|^2)\delta^{ij}- \Delta_B^{ij}) a^j = -i\omega g \varphi B_b^j, \\
        & (\omega^2 - |\Vec{k}|^2 -m^2) \varphi = i\omega g \Vec{a}\cdot \Vec{B}_b ,
    \end{split}
\end{equation}
with the auxiliary condition $\Vec{k}\cdot \Vec{B}_b=0$. We can see that only the component of the 3-vector potential parallel to the external magnetic field contributes to the mixing terms. By redefining the 3-vector potential in the form $\Vec{a}\rightarrow -i\Vec{a}$, we can make the coupling terms real. In matrix form, the field equations can be written as:
\begin{equation}
\label{MatrixFieldEquations}
    \left[(\omega^2 - |\Vec{k}|^2)\textbf{1}_3-2\omega\begin{pmatrix}
        \Delta_{\perp} & n_R & 0 \\
        n_R & \Delta_{||} & \Delta_M \\
        0 & \Delta_M & \Delta_a
    \end{pmatrix}\right] \begin{pmatrix}
        a_{\perp} \\
        a_{||} \\
        \varphi
    \end{pmatrix}= 0_{3 \times 1}.
\end{equation}
where $a_{||}$ and $a_{\perp}$ denote the components of the electromagnetic field parallel and perpendicular to the external magnetic field, respectively, and we define the quantities:
\begin{equation}
    \Delta_{\perp}=\frac{(\Delta_B)_\perp+\omega_p^2}{2\omega}, \ \ \ \Delta_{||}=\frac{\omega_p^2}{2\omega}, \ \ \ \Delta_a = \frac{m^2}{2\omega}, \ \ \ \Delta_M=\frac{gB_b}{2},
\end{equation}
where $(\Delta_B)_{\perp}$ represents the effects of the birefringence of the magnetized plasma medium on the 3-vector potential components perpendicular to the external magnetic field and $n_R$ represents the rotation of the polarization vector components due to the Faraday effect. 

Written this way, the field equations \eqref{MatrixFieldEquations} highlight a very interesting fact: the oscillating fields (photon and axion) are not simultaneous eigenstates of energy and 3-momentum. Another notable aspect is that only the component of the 3-vector potential parallel to the external magnetic field couples with the axion. Choosing energy eigenstates as a basis, and in the high-energy regime $\omega \ >> \ \omega_p$, we approximate the dispersion relation as $\omega \approx |\Vec{k}|$ and define $\Vec{k}=k_z \hat{z}$. Note that, in the high-frequency limit, we can neglect the effects of Faraday rotation. 

In this way, we can linearize the kinetic operator in the form:
\begin{equation}
\label{Linearization}
    (\omega^2 - |\Vec{k}|^2) = (\underbrace{\omega + k_z}_{=2\omega})(\omega + i\partial_z) \approx 2\omega (\omega + i\partial_z).
\end{equation}

\noindent With these approximations, we can express \eqref{MatrixFieldEquations} as:

\begin{equation}
\label{MatrixFieldEquations2}
    \left[(\omega + i\partial_z)\textbf{1}_3-\begin{pmatrix}
        \Delta_{\perp} & 0 & 0 \\
        0 & \Delta_{||} & \Delta_M \\
        0 & \Delta_M & \Delta_a
    \end{pmatrix}\right] \begin{pmatrix}
        a_{\perp} \\
        a_{||} \\
        \varphi
    \end{pmatrix}= 0_{3 \times 1}.
\end{equation}
where we return the 3-momentum $k_z$ to configuration space since the states are no longer 3-momentum eigenstates.

\subsection{Diagonalization and Mixing Matrix}

The non-diagonal sector of \eqref{MatrixFieldEquations2} can be written in a "Schrödinger-like" form:
\begin{equation}
\label{SchrodingerLikeEquation}
    -i\partial_z \begin{pmatrix}
        a_{||} \\
        \varphi
    \end{pmatrix} = 
    \underbrace{\begin{pmatrix}
    \Delta_{||} & \Delta_M \\
    \Delta_M & \Delta_a
    \end{pmatrix}}_{=H_I} \begin{pmatrix}
        a_{||} \\
        \varphi
    \end{pmatrix}. 
\end{equation}
The matrix $H_I$ in expression \eqref{SchrodingerLikeEquation} can be interpreted as a Hamiltonian matrix in 3-momentum space, as $-i\partial_z$ is a 3-momentum operator. The eigenvalues of $H_I$:
\begin{equation}
    \lambda_\pm = \frac{\Delta_{||}+\Delta_a}{2} \pm \sqrt{\frac{(\Delta_{||}-\Delta_a)^2}{4}+ \Delta_M}= \frac{\Delta_{||}+\Delta_a}{2} \pm \Delta_{osc}.
\end{equation}
Since the matrix $H_I$ is symmetric, it can be diagonalized by a $SO(2)$ matrix. Thus, the diagonal states are given by:
\begin{equation}
\label{DiagonalizationOfStates}
    \begin{pmatrix}
        a'_{||} \\
        \varphi'
    \end{pmatrix} = \underbrace{\begin{pmatrix}
        \cos \theta & \sin{\theta} \\
        -\sin{\theta} & \cos{\theta}
    \end{pmatrix}}_{=P}\begin{pmatrix}
        a_{||} \\
        \varphi
    \end{pmatrix}.  
\end{equation}
As the matrix P is a similarity transformation between the matrix $H_I$ in non-diagonal and diagonal form, i.e.:
\begin{equation}
    P^{-1}H_I P= H_D, \ \ \ \text{where} \ \ \ H_D= diag(\lambda^+, \lambda_-),
\end{equation}
we can show that the mixing angle $\theta$ is given by:
\begin{equation}
\label{MixingAngle}
    \sin{2\theta} = \frac{\Delta_M}{\Delta_{osc}}.
\end{equation}

With the diagonalization of $H_I$ complete—that is, with its eigenstates and eigenvalues determined—we have all the necessary material to derive the mixing matrix. From \eqref{SchrodingerLikeEquation}, it follows that $H_I$ is the generator of translations in the z-direction since its eigenvalues are 3-momenta $k_z$. Thus, we can write the translation operator z as:
\begin{equation}
    U(z) = e^{-iH_I z}.
\end{equation}

Now consider a beam of particles (photons or axions) in a region where there is an external magnetic field $\Vec{B}_b$. The eigenstates of $H_I$ after the particle beam travels a distance z:
\begin{equation}
    \begin{pmatrix}
        a'_{||}(z) \\
        \varphi'(z)
    \end{pmatrix}=U \begin{pmatrix}
        a'_{||}(0) \\
        \varphi'(0)
    \end{pmatrix}= \begin{pmatrix}
        e^{-i\lambda_+ z} & 0 \\
        0 &  e^{-i\lambda_- z}
    \end{pmatrix}\begin{pmatrix}
        a'_{||}(0) \\
        \varphi'(0)
    \end{pmatrix}.
\end{equation}
Returning to the non-diagonal basis, we have: 
\begin{equation}
    \begin{pmatrix}
        a_{||}(z) \\
        \varphi(z)
    \end{pmatrix}=\underbrace{P^{-1}UP}_{=M} \begin{pmatrix}
        a_{||}(0) \\
        \varphi(0)
    \end{pmatrix},
\end{equation}
where the mixing matrix M is given by:
\begin{equation}
\label{MixingMatrix}
    M=\begin{pmatrix}
        \cos^2{(\theta)}e^{-i\lambda_+z} + \sin^2{(\theta)}e^{-i\lambda_-z}, & -i\sin{(2\theta)}e^{i(\Delta_{||}+\Delta_a)z/2}\sin{(\Delta_{osc}z)} \\
        -i\sin{(2\theta)}e^{i(\Delta_{||}+\Delta_a)z/2}\sin{(\Delta_{osc}z)}, & \cos^2{(\theta)}e^{-i\lambda_-z} + \sin^2{(\theta)}e^{-i\lambda_+z}
    \end{pmatrix}.
\end{equation}

The matrix \eqref{MixingMatrix} teaches us something extremely curious. It indicates that the axion and photon fields, after traveling a distance z in the background magnetic field, are a linear combination of the fields. Therefore, the off-diagonal elements give us the photon/axion transition probability:
\begin{equation}
\label{AxionPhotonProbability}
    \boxed{P_{a_{||} \rightarrow \varphi} = |M_{12}|^2=|\bra{a_{||}(0)}U(z)\ket{\varphi(0)}|^2 = \sin^2{(2\theta)}\sin^2{(\Delta_{osc}z)}}
\end{equation}
Note that, as the mixing matrix \eqref{MixingMatrix} is symmetric, the axion/photon transition probability is the same. The parameter $\Delta_{osc}$ has dimensions of $\text{length}^{-1}$ and can be used to estimate the scale at which the oscillation can be observed through the so-called oscillation length $l=2\pi/\Delta_{osc}$.

\subsection{Axion-Photon Mixing via Linearized Propagators}

There is an alternative method, more direct but less explicit, to obtain the axion/photon transition probability through linearized propagators, i.e., correlation functions in 3-momentum space. To do so, we need to study the Lagrangian \eqref{AxionEDLagrangian} in 4-momentum space. First, let us rewrite the Lagrangian \eqref{BackgroundApproximationLagrangian}, in the background field approximation, in the form:
\begin{equation}
	\begin{split}
    		\mathcal{L}_{AX}(x) &= \frac{1}{2}a^\mu(\Box \eta_{\mu\nu}-			\partial_\mu \partial_\nu)a^\nu-J^\mu a_\mu -\frac{1}{2}				\varphi(\Box+m^2)\varphi \\   
    		&+\frac{g}{2} \varphi \partial_\mu a_\nu \Tilde{F}_B^{\mu\nu}-\frac{g}{2}\partial_\mu \varphi \Tilde{F}_B^{\mu\nu}a_\nu
    	\end{split}
\end{equation}
In 4-momentum space:
\begin{equation}
\label{AxionEDLagrangian4Momentum}
	\begin{split}
    		\mathcal{L}_{AX}(k) &= \frac{1}{2}a^\mu (-k^2 \eta_{\mu\nu}			+k_\mu k_\nu)a^\mu-J^\mu a_\nu -\frac{1}{2}\varphi (-k^2+m^2)			\varphi\\
    		&-\frac{ig}{2} \varphi k_\mu a_\nu \Tilde{F}_B^{\mu\nu}+\frac{ig}{2}k_\mu \varphi \Tilde{F}_B^{\mu\nu}a_\nu.
    	\end{split}
\end{equation}
Again, assuming the same conditions imposed on \eqref{FourierFieldEquations} ($\Vec{k}\cdot \Vec{a}=0$ and $\Vec{a}\rightarrow i\Vec{a}$) and writing the 3-vector potential as $\Vec{a}=a_\perp + a_{||}$, we can express the Lagrangian \eqref{AxionEDLagrangian4Momentum} as:
\begin{equation}
    \mathcal{L}_{AX}(k) =  \frac{(k^2)}{2}a_\perp^2 + \frac{1}{2}\begin{pmatrix}
        a_{||} & \varphi
    \end{pmatrix}
    \underbrace{\begin{pmatrix}
        \omega^2-\omega_p^2 - (\Delta_B)_{||}-|\Vec{k}|^2 & \omega gB_b  \\
        \omega gB_b & \omega^2-|\Vec{k}|^2-m_a^2
    \end{pmatrix}}_{=K_c}\begin{pmatrix}
        a_{||} \\
        \varphi
    \end{pmatrix}
\end{equation}

The kinetic matrix $K_c$ is the Lagrangian version of the matrices \eqref{MatrixFieldEquations}. They demonstrate once again that the fields are not simultaneously energy and 3-momentum eigenstates. With the didactic goal of showing that both methods lead to the same transition probability, we choose energy eigenstates here and linearize the kinetic operators analogously to \eqref{Linearization} (we do not return the 3-momentum vector to configuration space here because, at the Lagrangian level, the fields do not obey the physical dispersion relations):
\begin{equation}
    \mathcal{L}_{AX}(k) = \frac{(-k^2)}{2}\phi^2 + \frac{(k^2)}{2}a_\perp^2 + \omega^2(a_\perp^2 + \varphi^2) +\omega \begin{pmatrix}
        a_{||} & \varphi
    \end{pmatrix}
    \underbrace{\begin{pmatrix}
        -k_z-\Delta_{||} & \Delta_M \\
        \Delta_M & -k_z-\Delta_a
    \end{pmatrix}}_{=G^{-1}(\omega,k_z)}\begin{pmatrix}
        a_{||} \\
        \varphi
    \end{pmatrix}. 
\end{equation}

The matrix $G^{-1}(\omega, k_z)$ has wave operators in $k_z$ momentum space as its elements. The inverse matrix is formed by linearized propagators, i.e., correlation functions in $k_z$ momentum space:
\begin{equation}
\label{CorrelationMatrix}
    M(\omega,k_z)\equiv G(\omega, k_z) = \frac{1}{k_z^2 + (\Delta_a+\Delta_{||})k_z +\Delta_a \Delta_{||}  - \Delta_M^2}\begin{pmatrix}
        -k_z-\Delta_a & -\Delta_M \\
        -\Delta_M & -k_z-\Delta_{||}
    \end{pmatrix}
\end{equation}
The matrix \eqref{CorrelationMatrix} contains the transition amplitudes in $k_z$ momentum space. Therefore, if we perform the inverse Fourier transform and return to configuration space, we obtain the mixing matrix \eqref{MixingMatrix} again. Thus, the axion/photon transition amplitude:
\begin{equation}
\label{MixingCorrelation}
    \begin{split}
    G_{a_{||}\rightarrow \varphi}(z, \omega) &= -\Delta_M \int \frac{dk_z}{2\pi} \frac{e^{ik_z z}}{k_z^2 + (\Delta_a+\Delta_{||})k_z +\Delta_a \Delta_{||}  - \Delta_M^2} \\
    &= -\frac{\Delta_M}{(4\pi)\Delta_{osc}}\int dk \ e^{ik_z z}\left[\frac{1}{k_z+\lambda_+}-\frac{1}{k_z + \lambda_-}\right] =\frac{-i\Delta_M}{2\Delta_{osc}} \left[e^{-i\lambda_+ z}-e^{-i\lambda_- z}\right],\\
    &= -\frac{\Delta_M}{\Delta_{osc}}e^{-i(\Delta_{||}+\Delta_a)z/2} \sin{(\Delta_{osc}z)}= -iM_{12}.
    \end{split}
\end{equation}

We can see, with the help of relation \eqref{MixingAngle}, that the correlation function \eqref{MixingCorrelation} is equal to the off-diagonal element of matrix \eqref{MixingMatrix}, up to a phase (which does not contribute to the transition probability). Therefore, the squared modulus of the correlation function \eqref{MixingCorrelation} gives us the axion/photon transition probability:
\begin{equation}
\label{probability}
    \boxed{P_{a_{||}\rightarrow \varphi}(z)=|G_{a_{||}\rightarrow \varphi}(\omega,z)|^2 = \left(\frac{\Delta_M}{\Delta_{osc}}\right)^2 \sin^2{(\Delta_{osc}z)}}
\end{equation}
 which is the same result obtained in \eqref{AxionPhotonProbability}. 
 
 With the result \eqref{probability} established, we concluded the demonstration of the Primakoff effect. However, as can be seen from section \ref{SunProfile}, the Sun's magnetic field is not homogeneous through the stellar medium, which requires an new approach to consider the effects of the inhomogeneity of the background magnetic field. We present this method in the next section.

\section{The WKB Approximation in the Presence of Inhomogeneous Background Fields}
\label{WKBsolutionInhomogeneousMagneticField}

Our framework so far considers the Primakoff process in the context of a constant and homogeneous background magnetic field. However, this is not the case for a solar magnetic field, as can be seen from figure \ref{SunsMagneticField}. The literature considers only the effects of axion-photon transitions in the presence of constant magnetic field \cite{AxionLikeProduction}. In this section,  we would like to consider the phenomenology discussed so far in the case of a constant inhomogeneous background magnetic fields. In order to consider the effects of the inhomogeneity of the stars's magnetic fields, we turn back to equation \eqref{WaveEqInPlasmaMagneticField}, in configuration space,
\begin{equation}
\label{WaveEquation1}
    (\varepsilon(r)\partial_t^2 - \nabla^2)\vec{E}(\vec{r},t)=0,
\end{equation}
but now we propose the WKB approximation solution \cite{liu1973theory,tracy2014ray,weinberg2013lectures},
\begin{equation}
    \vec{E}(\vec{r},t)= \vec{E}_0(r)e^{-i\omega t + i\omega\psi(r)},
\end{equation}
with the definition that $\vec{n}(r)=\vec{k}(r)/\omega=\vec{\nabla}\psi(r)$.
Substituting into the equation \eqref{WaveEquation1}, we find:
\begin{equation}
    \begin{split}
        &-\omega^2 \varepsilon(r)\vec{E}_0 -\left[\nabla^2 \vec{E}_0 +2i\omega\vec{\nabla}\psi \vdot \vec{\nabla}(\vec{E}_0)+i\omega  \nabla^2\psi (\vec{E}_0)-(\vec{\nabla}\psi)^2\vec{E}_0 \right]=0 \\
        & \left[\varepsilon(r) - (\vec{\nabla} \psi)^2\right] \vec{E}_0 -\frac{1}{\omega}\left[2i\vec{\nabla}\psi \vdot \vec{\nabla}(\vec{E}_0)+i\nabla^2 \psi (\vec{E}_0)\right] +\frac{\nabla^2 \vec{E}_0}{\omega^2} = 0.
    \end{split}
\end{equation}
Assuming high frequency regime (for our considerations, we need $\omega >> \omega_p,\omega_B$), we find, in order zero in the ($1/\omega$) parameter, that
\begin{equation}
    (\vec{\nabla}\psi)^2\delta^{ij} = \varepsilon^{ij}(r)=(n^2)^{ij}(r)= \frac{|\vec{k}|^2(r)\delta^{ij}}{\omega^2},
\end{equation}
i.e. gives us the local dispersion relation, for a given polarization mode, with $\vec{\nabla}\psi=\vec{n}$. Therefore, the eikonal phase is given by
\begin{equation}
    \psi(r)= \int_{0}^r \vec{n}(r')\vdot d\vec{r'}.
\end{equation}

The term in first order in $1/\omega$ gives us the amplitude:
\begin{equation} 
    2\vec{n}\vdot \vec{\nabla} \vec{E}_0(r) = - (\vec{\nabla}\vdot\vec{n}) \vec{E}_0(r). 
\end{equation}
  Using the notation $\vec{n}_\alpha=n_\alpha \hat{s}$, with $\alpha$ indicating the polarization mode, we find the amplitude as
\begin{equation}
    \vec{E}_0(r) = \frac{E_0 \vec{e}_\alpha}{\sqrt{n_\alpha(r)}},
\end{equation}
with $A_0$ a real number. Therefore, we can write the WKB solution as,
\begin{equation}
\label{EikonalAnsatz}
    \boxed{\vec{E}_\alpha(r)=\frac{E_0 \vec{e}_\alpha} {\sqrt{n_\alpha(r)}} \exp{i\omega\int^{r}_{0} \vec{n}_\alpha(r')\vdot d\vec{r'}-i\omega t}}
\end{equation}
In the next section, we consider the case of circular polarization mode and proceed to demonstrate the Faraday effect in an inhomogeneous field.

\subsection{Faraday Effect in an Inhomogeneous Field}

Considering the dispersion relation for circularly polarized mode \eqref{PlasmaDispersion}, in the limit of high frequencies ($\omega >> \omega_p, \omega_B$), 

\begin{equation}
    n_\pm(r) =\frac{|\vec{k}_\pm|(r)}{\omega} = 1 - \frac{\omega_p^2(r)}{2\omega^3}(\omega \pm \omega_B(r)), 
\end{equation}
and the associated eikonal phase is given as
\begin{equation}
    \psi_\pm = \int_0^r n_\pm (r')dr'= \int_0^r \left[1 - \frac{\omega_p^2(r')}{2\omega^3}(\omega \pm \omega_B(r'))\right]dr'=\psi_0 \pm \Delta\psi,
\end{equation}
the linearly polarized wave can be written as a linear combination of both circular modes:
\begin{equation}
    \begin{split}
        \vec{E} &= \frac{A(r)}{\sqrt{2}}\left(\vec{e}_+ e^{-i\omega(t-\psi_+)}+\vec{e}_- e^{-i\omega(t-\psi_-)}\right) \\
        &= \frac{A(r)}{\sqrt{2}} e^{-i\omega(t-\psi_0)}(\vec{e}_+e^{i\omega\Delta\psi}+\vec{e}_-e^{-i\omega\Delta\psi})\\
        &= \frac{A(r)}{\sqrt{2}} e^{-i\omega(t-\psi_0)} 2\left[\cos{(\omega \Delta \psi})\vec{e}_1 - \sin{(\omega \Delta \psi)\vec{e}_2}\right]
    \end{split}
\end{equation}
The real part of the wave is
\begin{equation}
    \Re{\vec{E}}= \frac{A(r)}{\sqrt{2}} \cos{(\omega(t-\psi_0))}\left[\cos{(\omega \Delta \psi})\vec{e}_1 - \sin{(\omega \Delta \psi)\vec{e}_2}\right],
\end{equation}
and the associated birefringent phase is given by
\begin{equation}
    \boxed{\beta(r)=\tan^{-1}{\frac{E_2}{E_1}} = -\omega \Delta \psi(r) = \int_0^r \frac{\omega_p^2(r')}{2\omega^2}\omega_B (r') dr'} 
\end{equation}
\subsection{Kinetic Mixing in the Presence of Inhomogeneous Fields}

We wish to generalize the Primakoff effect to encompass the inhomogeneity of the stellar magnetic field. Returning to the field equations \eqref{LorenzGaugeFieldEquations}, with $\vec{\nabla}\vdot \vec{a}=0$, i.e.
\begin{equation}
\label{equation1}
    \begin{split}
        &(\Box - \omega_p^2(r)) a_{||}(\vec{r},t) = -g \partial_t \varphi(\vec{r},t) B_b(r)\\
        &(\Box - m^2) \varphi(\vec{r},t) = g B_b(r) \partial_t a_{||},
    \end{split}
\end{equation}
we now propose that the fields have the form:
\begin{equation}
    \begin{cases}
        &a_{||} (\vec{r},t)= \frac{a_{||}(r)}{\sqrt{k(r)}} e^{i\int k(r)dr -i\omega t}, \\
        &\varphi (\vec{r},t)= \frac{\varphi(r)}{\sqrt{k(r)}} e^{i\int k(r)dr-i\omega t}.
    \end{cases} 
\end{equation}
Substituting in  \eqref{equation1}, we find:

\begin{equation}
    \begin{split}
        &(\omega^2 - |\vec{k}|(r) + \omega^2_p(r))a_{||}(r)-2\omega |\vec{k}|(r)i\partial_r a_{||}(r)= \omega g B_b(r)\varphi(r),\\
        &(\omega^2 - |\vec{k}|(r) + m^2)\varphi(r)-2\omega |\vec{k}|(r)i\partial_r \varphi(r)= \omega g B_b(r)a_{||}(r).\\
    \end{split}
\end{equation}

Next, we assume the high frequency dispersion relation $|\vec{k}|(r)\sim \omega$. Then, we can write the field equations as
\begin{equation}
\label{InhomogeneousMixingEq}
    -i\partial_r \begin{pmatrix}
        a_{||}(r) \\
        \varphi(r)
    \end{pmatrix}=\underbrace{\begin{pmatrix}
        \frac{\omega_p^2(r)}{2\omega} & \frac{gB_b(r)}{2} \\
        \frac{gB_b(r)}{2} & \frac{m^2}{2\omega}
    \end{pmatrix}}_{=H_I(r)}\begin{pmatrix}
        a_{||}(r) \\
        \varphi(r)
    \end{pmatrix},
\end{equation}
which defines a local interaction mixing Hamiltonian $H_I(r)$. Therefore, $H_I(r)$ is a generator of a local translation. We could think of \eqref{InhomogeneousMixingEq} as a "Schröedinger-like" equation,
\begin{equation}
    \begin{split}
        &-i\partial_r \vb{\Psi}_i (r) = (H_I(r))_{ij}\vb{\Psi}_j (r), \\
         &-i\partial_r U_{ik} (r,) \vb{\Psi}_k (0) = (H_I(r))_{ij}U_{jk}(r)\vb{\Psi}_k (0)
    \end{split}
\end{equation}
which has the formal solution for $U(r)$ as
\begin{equation}
    U_{ij}(r) = R \exp{i\int_{0}^r (H_I)_{ij}(r') dr'},
\end{equation}
where the operator "R" denotes "spatial ordering". Once again, we can diagonalize the local Hamiltonian in the same manner as in \eqref{DiagonalizationOfStates}, but now we find local eigenvalues, i.e.
\begin{equation}
    \lambda_\pm(r) = \frac{\Delta_{||}(r)+\Delta_a}{2} \pm \sqrt{\frac{(\Delta_{||}(r)-\Delta_a)^2}{4}+\Delta_M(r)}=\frac{\Delta_{||}(r)+\Delta_a}{2} \pm \Delta_{osc}(r),
\end{equation}
and the mixing angle given by,
\begin{equation}
    \sin{2\theta(r)} = \frac{\Delta_M(r)}{\Delta_{osc}(r)}.
\end{equation}
Space translating the diagonal states, we find:
\begin{equation}
    \begin{pmatrix}
        a_{||}'(r) \\
        \varphi'(r)
    \end{pmatrix} = e^{i\alpha(r)}\begin{pmatrix}
        e^{i\chi(r)} & 0 \\
        0 & e^{-i\chi(r)}
    \end{pmatrix} \begin{pmatrix}
        a_{||}'(0) \\
        \varphi'(0)
    \end{pmatrix},
\end{equation}
where we defined the integrated phase,
\begin{equation}
    \alpha(r)=\int^r_{0} \frac{\Delta_{||}(r')+ \Delta_a}{2}dr', \ \ \ \textrm{and} \ \ \ \chi(r) = \int^r_{0} \sqrt{\frac{(\Delta_{||}(r')-\Delta_a)^2}{4}+ \Delta_M(r')}dr'.
\end{equation}
In the interaction base, we find:
\begin{equation}
    \begin{pmatrix}
        a_{||}(r) \\
        \varphi(r)
    \end{pmatrix} = M(r,0)  \begin{pmatrix}
        a_{||}(0) \\
        \varphi(0)
    \end{pmatrix},
\end{equation}
with
\begin{equation}
    M(r,r_0) = e^{i\alpha(r)}\begin{pmatrix}
        \cos{\Delta_{\theta -}\cos{\chi(r)}+ i\cos{\Delta_{\theta +}}\sin{\chi(r)}}; & i\sin{\Delta_{\theta +}}\sin{\chi(r)} - \sin{\Delta_{\theta -}}\cos{\chi(r)} \\
        i\sin{\Delta_{\theta +}}\sin{\chi(r)} + \sin{\Delta_{\theta -}}\cos{\chi(r)}; & \cos{\Delta_{\theta -}}\cos{\chi(r)} - \cos{\Delta_{\theta +}}i\sin{\chi(r)}
    \end{pmatrix},
\end{equation}
where we defined the quantities:
\begin{equation}
    \Delta_{\theta \pm} = \theta(r) \pm \theta(0),
\end{equation}
with $\theta(r)=2^{-1}\sin^{-1}{[\Delta_M(r)/\Delta_{osc}(r)]}$. The modulus squared of the off-diagonal terms gives us the transition amplitude:
\begin{equation}
\label{ProbInhomogeneous}
    \boxed{P_{\gamma \rightarrow a}(r) = \sin^2{\Delta_{\theta +}}(r) \sin^2{\chi(r)} + \sin^2{\Delta_{\theta -}}(r)\cos^2{\chi(r)}}
\end{equation}

As can be seen from \eqref{ProbInhomogeneous}, if the mixing angle $\theta$ is constant through the star, we have $\Delta_{\theta_-}=0$, $\Delta_{\theta_+}=2\theta$ and $\chi(r)=\Delta_{osc}r$, and we recover the result \eqref{AxionPhotonProbability} deduced early. The probability transition is modulated by the mixing angles in different points in the star, which is a direct consequence of the inhomogeneity of the background magnetic field.  Our formalism so far established the 1-particle axion-photon transitions. In the following section, we propose a density matrix formalism to consider the effects of mixing transitions on ensembles of thermal photons and axions.

\section{Flux of Axions Produced in the Sun}

We can calculate the production rate of axions/photons via some thermal mechanism using the density operator formalism and the so-called quantum Liouville equation \cite{redondo2013solar}. Through this observable, we can investigate the consequences of axion production by stellar plasma on the structure of stellar objects, such as the Sun, or in the physics of degenerate stars, such as white dwarfs \cite{raffelt1996stars}, using the so-called "energy-loss argument". The matrix $H_I$ can be rewritten as

\begin{equation}
H_I=\begin{pmatrix}\Delta_{||} & \Delta_M \\
\Delta_M & \Delta_a
\end{pmatrix}= \frac{\Delta_{||} + \Delta_a}{2} \textbf{1}_2 + \begin{pmatrix}
+\Delta \omega & \Delta_M \\
\Delta_M & -\Delta \omega
\end{pmatrix},
\end{equation}
where $\Delta \omega=(\Delta_{||}-\Delta_a)/2$. Here, we consider an ensemble (a set of states, with their respective probabilities, that the system can occupy at a given temperature) of axions/photons interacting with a background magnetic field without 3-momentum scattering. The density matrix at thermal equilibrium is:

\begin{equation}
\rho_T= \begin{pmatrix}f_T & 0 \\
0 & 0
\end{pmatrix},
\end{equation}
where $f_T= (e^{\omega /T}-1)^{-1}$ is the Bose-Einstein distribution, which gives us the occupation of quantum states at a given temperature $T$ (here we consider $\hbar=k_b=1$). Since the magnetic background can interact with the photon and axion fields by absorbing and emitting energy quanta, the time evolution of the density operator is given by the quantum Liouville equation:

\begin{equation}
\dot{\rho}= -[H_I, \rho] + \frac{1}{2}{G_{prod},1+\rho}-\frac{1}{2}{G_{abs},\rho},
\end{equation}
where we have defined the phenomenological particle absorption/emission matrices,

\begin{equation}
G_{\textrm{prod/abs}}=\begin{pmatrix}\Gamma_{\textrm{prod/abs}} & 0 \\
0 & 0
\end{pmatrix},
\end{equation}
with $\Gamma_{prod/abs}$ being the absorption or emission rate of axions/photons with 3-momentum $\Vec{k}$. At thermal equilibrium, they obey $\Gamma_{prod}=e^{-\omega T}\Gamma_{abs}$.At thermal equilibrium, we have $\dot{\rho}=0$ and the final states are not excited. However, by introducing a small perturbation around thermal equilibrium:

\begin{equation}
    \rho=\rho_T + \delta \rho =\begin{pmatrix}f_T & 0 \\ 
    0 & 0
    \end{pmatrix} + \begin{pmatrix} n_0 & g \\ 
    g^* & n_e
\end{pmatrix}.
\end{equation}
The time evolution equation becomes:

\begin{equation}
\dot{\rho} = -i[H_I,\rho] -\frac{1}{2}{G, \delta \rho}.
\end{equation}
The matrix $G=\textrm{diag}(\Gamma,0)$ contains the "damping" rate $\Gamma=\Gamma_{abs}-\Gamma_{prod}=(1-e^{-\omega T})\Gamma_{abs}$. The equations take the form:

\begin{equation}
    \begin{pmatrix}
        \dot{n_0} & \dot{g} \\
        \dot{g^*} & \dot{n_e}
    \end{pmatrix}=\begin{pmatrix}
        -2\Delta_M\Im(g) - \Gamma n_0 & -g(i\Delta \omega + \Gamma/2)+i\Delta_M(f_T + n_0 - n_e) \\
        -g^*(-i\Delta \omega + \Gamma/2)-i\Delta_M(f_T + n_0 - n_e) & 2\Delta_M\Im(g)
    \end{pmatrix}.
\end{equation}
Assuming that the excitations do not deviate significantly from thermal equilibrium (after all, $\delta \rho$ is merely a small perturbation), i.e., $n_0,n_e \ll f_T$:

\begin{equation}
\label{eqDiferencial1}
    \dot{g}= -g(i\Delta \omega + \Gamma/2)+i\Delta_M f_T.
\end{equation}
Assuming the initial condition $g(0)=0$, differential equation \eqref{eqDiferencial1} has the following decoherence solution:

\begin{equation}
\label{SolucaoDecoerencia}
    g(t)= \frac{1-e^{-i(\Delta \omega+\Gamma/2)t}}{\Delta \omega-i\Gamma/2}\Delta_M f_T.
\end{equation}
After an initial transient phase, solution \eqref{SolucaoDecoerencia} reaches the steady state:

\begin{equation}
    g(\infty)= \frac{\Delta \omega+i\Gamma/2}{(\Delta \omega)^2 +\Gamma^2/4}\Delta_M f_T.
\end{equation}
Inserting the steady-state decoherence solution into the equation for $\dot{n}_e$, we finally obtain the production rate of axions/photons in the presence of an external magnetic field:

\begin{equation}
\label{TaxaDeProducao}
    \boxed{\dot{n_e} = \frac{\Gamma \Delta_M^2}{(\Delta_{||}-\Delta_a)^2+\Gamma^2/4}\frac{1}{e^{\omega/T}-1}}.
\end{equation}

Note that this production rate is valid both under resonant conditions ($\Delta_{\vert{}\vert{}}=\Delta_a$) and off-resonance, and the only approximation made here is that the couplings $\Delta_M$ are much smaller than the separation of the energy levels $\Delta \omega$.

\subsection{Flux of Axions Produced in the Sun}

In this section, we will discuss how we can constrain the parameter space of Axion Electrodynamics using solar stellar models. Using the thermal production rate of axions, we can calculate the total luminosity produced by axions and contrast it with solar data. The axion flux reaching Earth is given by \cite{AxionLikeProduction}:

\begin{equation}
    \label{FluxoAxionsTerra}
    \frac{dN_a}{dt}=\frac{g}{4\pi D_\odot} \int_\odot d^3r \frac{d^3k}{(2\pi)^3}\delta(\omega - |\vec{k}|(r)) \Gamma_a^{prod},
\end{equation}
where $D_\odot=1.49 \times 10^{11} \text{ m}$ is the Earth-Sun distance, $\Gamma^{prod}_a$ is the production rate of axions via the Primakoff effect as expressed in \eqref{TaxaDeProducao}, $g$ is the number of photon polarization states ($g=1$ for longitudinal modes and $g=2$ for transverse modes), and the integration is performed over all $3$-space $\vec{k}$ , assuming the high-energy dispersion relation $\omega\approx |\vec{k}|$ (which justify the delta function $\delta(\omega - |\vec{k}|)$), and over the solar volume. The latter requires using knowledge from the standard solar model \cite{iben2012stellar} and the profile of the Sun's magnetic field \cite{gough1990effect}.

A final ingredient necessary to integrate \eqref{FluxoAxionsTerra} is the photon production rate in the solar interior. Detailed calculations show that, under thermal equilibrium conditions, $\Gamma^{prod}_a=\Gamma^{abs}_a e^{-\omega/T}$, where $\Gamma^{abs}_a$ is the photon absorption rate in the stellar interior due to photon interaction physics. The latter, in natural units, is inversely proportional to the photon mean free path. Thus, $\Gamma^{abs}_a$ depends on the solar electron density and the cross section of the physical processes related to the opacity of the star's interior. In the case of the Sun, at low and medium energies, the predominant source of opacity is inverse bremsstrahlung scattering (called free-free absorption) and Thomson scattering at high energies. For transverse modes, the absorption rate is given by:

\begin{equation}
    \label{EmissionRatePhotons}
    \Gamma_{abs}(r,\omega) = \frac{64 \pi^3 \alpha^3}{3 m_e^{3/2}\sqrt{2\pi T(r)}}\frac{n_e(r)}{\omega^3} \sum_j Z_j^2 n_j F\left(\frac{\omega}{T(r)}\right)+ \frac{8\pi\alpha^2n_e(r)}{3m_e^2};
\end{equation}
where $n_e(r)$ is the free electron density and $n_j$ is the ion density of the $j$-th species with proton number $Z_j$. Note that the particle densities are proportional to the star density $\rho(r)$, which together with the temperature $T(r)$ are given by the solar profile \eqref{SunProfile}.  

The function $F(\omega/T(r))$,
\begin{equation}
    F\left(\frac{\omega}{T(r)}\right)= K_0 \left(\frac{\omega}{2T(r)}\right) \sinh{\left(\frac{\omega}{2T(r)}\right)};
\end{equation}
where $K_0(\omega/T(r))$ is a  Bessel function of the second kind, accounts for the quantum effects from integrating the cross section of electron scattering off an ion nucleus in the presence of a photon, where the nuclear potential is modeled by a Coulomb potential modulated by a negative exponential that incorporates the charge shielding effects of the nucleus as the relative distance from the electron to the nucleus changes (also known as a "Yukawa-type" potential). For a detailed and didactic review of this subject, we recommend reading \cite{iben2012stellar}. 

Finally, it is mandatory to emphasize that, in approximation \eqref{EmissionRatePhotons}, other processes contributing to stellar opacity—such as bound-bound and free-bound transitions, Pauli blocking, and Sommerfeld enhancement—were neglected, although the contribution of these processes is small given the typical energy regime for stars at the Sun's temperature. We also treated the Sun as a fully ionized star in the calculation of the electron density. This approximation fails near the boundaries of the star, requiring the solution of the Saha equation to obtain a more accurate description of ionization. Since the largest contribution to the plasma frequency \eqref{PlasmaFrequency} and the photon production rate from \eqref{EmissionRatePhotons} originates from the stellar core region, the errors near the solar boundary remain small.

\subsection{Bounds on Axion's Coupling from Sun's Physics}

Since the axions don't interact electromagnetically, it constitutes a source of energy loss for the sun. Based on this feature, we wish to impose bounds on the axion's parameter space in the presence of an inhomogeneous background magnetic field. Assuming relativistic states $\omega \approx |\va{k}|$, we can insert a delta function $\delta(|\va{k}|- \omega)$ in the photon's 3-momentum integral \eqref{FluxoAxionsTerra}. Therefore, the photon differential spectrum measured on Earth is:

\begin{equation}
\label{DifferentialSpectrum}
    \dv{\Phi_a}{\omega} = \frac{1}{4\pi D^2_{\odot}}\frac{g^2}{ \pi} \int^{R_\odot}_0 dr \frac{\ r^2 B_b^2(r)}{\left(\frac{\omega_p^2(r)-m^2}{2\omega}\right)^2+ \Gamma(\omega,r)} \frac{\omega^2 \Gamma(\omega,r)}{e^{\omega/T(r)}-1},
\end{equation}
with $\Gamma(\omega,r)=(1-e^{-\omega/T(r)})\Gamma_{abs}(\omega,r)$ and $\Gamma_{abs}$ given by \eqref{EmissionRatePhotons}. For the axion mass, we consider the bounds imposed by supernovae \cite{burrows1989axions} ($m\sim0.02 \ eV$) . Integrating \eqref{DifferentialSpectrum} numerically, we obtain the following axion luminosity:
\begin{equation}
    \boxed{L_a= 8.28 \times 10^{28} g_0^2L_{\odot}},
\end{equation}
where $L_\odot=1.579 \times 10^{30} \ eV^2$ is the Sun's luminosity and we defined the dimensionless constant $g_0=g/(10^{-10}  GeV^{-1})$.
According to the so-called "energy loss argument", this energy depletion must be lower than our precision measurements of the sun's luminosity. The best constraints come from helioseismological and neutrino emission observations. Following \cite{krief2016solar}, we can impose a bound on the axion luminosity,

\begin{equation}
    L_a \leq 0.003  \ L_\odot,
\end{equation}
which implies the following bound on the axion's coupling constant,
\begin{equation}
\label{FermionicBound}
    \boxed{g \leq 1.903 \times 10^{-6} \ GeV^{-1}}
\end{equation}

As can be seen from figure \ref{EspacoDeParametrosAxion}, this bound is in the range of the parameter space of the CAST experiment, but it is far from the most limiting bound $g\sim 10^{-10} \ GeV^{-1}$. This is expected, since most bounds consider the case of constant homogeneous magnetic fields. However, our model is able to deal with the inhomogeneous background magnetic field, which makes our bounds more restrictive. 

\section{Conclusion and Final Remarks}
\label{Conclusion}

In this Section, we would like to highlight possible perspectives for future research coming from our present results. In the derivation of the Primakoff effect, we considered in \eqref{MatrixFieldEquations2}  the propagation of the photon's transverse mode and its subsequent mixing with the axion mode. Due to the presence of the stellar plasma, longitudinal photon modes are also present \cite{AxionLikeProduction} with an associated Primakoff effect. The contributions from the plasmon modes to the production of axions in the Sun's interior has already been considered in the literature \cite{redondo2013solar}, however in presence of a constant background magnetic field. The path followed here could be extended to consider the effects of the inhomogeneity of the background magnetic field on the production of axions from plasmon modes, and more restrictive bounds on the axion-photon coupling can de archive. 

We would like to emphasize that the new framework presented here could be used to study the effects induced by non-homogeneous background fields in other physical scenarios, for instance the QED in the presence of extreme background fields or even different electrodynamics \cite{sharma2026strong}.  The appearance of background fields alters the structure of the vacuum and our methods allow to predict effects from inhomogeneities present in the medium. This kind of scenario could be inspected in QASAR experiments \cite{Weinsheimer:2018ulb}. In shining-light-through-wall-type experiments, a collimated laser crosses a region with a strong magnetic field due to the presence of a superconducting LHC dipole. An optical barrier is introduced on the path, with an aligned CCD on the opposite side of the barrier. Photons modes transverse to the magnetic field could oscillate into ALPs and penetrate the optical barrier, since ALPs does not interact electromagnetically. On the other side, ALPs could oscillate again into photons and generate a signal at the CCD. Our methods could improve the predictions from these experiments by considering the inhomogeneity of the magnetic fields, and improve the imposed bounds on the ALPs parameter space.

In the astrophysical scene, our formalism can be used to improve the effects of ALP-photon oscillations on the observed spectra of Blazars. The latter is made of two collimated relativistic jets (quasars), originated from supermassive black holes, that are oriented in Earth's line of sight \cite{dingus2001highest,giorgio2025blazars}. The spectra of Blazars is composed by a broad range of energies, from radio wavelengths to photons with energies above $10 \ TeV$, and in particular the spectral distribution has two distinctive broad humps. One of them is present in the IR-UV band, originated from syncroton emission by relativistic electrons in the jets, and the second one is present in the gamma-ray energies. This high energy hump is still topic of discussion. It is expected that this gamma-ray photons are absorved by extragalatic background light (IR-optical-UV range) by the QED vacuum process $\gamma \gamma \rightarrow e^-e^+$. The presence of the excess of photons could be explained by the oscillation of the high-energy photons into ALPs, which does not interact electromagnetically. Our treatment of the ALP-photon mixing could improve the Blazars spectral by considering the effects of the inhomogeneity of the extragalatic background magnetic field, and even the measurements of the gamma-ray photons degree of linear polarization. 

Another interesting astrophysical context for the Primakoff effect is the prominent magnetic fields present in magnetars and pulsars \cite{gardner2025pulsar,chatterjee2019magnetic}. These are neutral stars with magnetic field up to $10^{14} \ T$, with the latter one being  a rotating neutron stars. These fields are much higher than Schwinger's critical field $B_{crit}=10^9 \ T$ \cite{brevik2025below}, which allows the possibility of polarization of the vacuum with the creation of electron-positron pairs. This polarization effect opens up the path to non-linear electrodynamics, as in the case of Euler-Heisemberg Electrodynamics \cite{kruglov2017remarks}. These non-linear effects generate modifications on the dispersion relations \cite{santos2024dispersion} that could amplify the luminosity loss from neutron stars through the Primakoff effect. Also, the birefringence created by the vacuum polarization and the ALP-photon oscillation could also be measured as signatures in the degree of linear polarization of the light arriving at Earth observatories \cite{bailey2010linear}.

\section*{Acknowledgments}

The presence work was developed with the support of the Brazilian Federal Agency for Support and Evaluation of Graduate Education (CAPES foundation) - financing code 001. The author expresses his gratitude to the Federal University of Rio de Janeiro for the hospitality and for providing an academic atmosphere that enables high-level research in Brazil. Thanks are also due to J. A. Helayël-Neto for my PhD Thesis supervision and fruitful discussions.

\bibliographystyle{plain}
\bibliography{referencias.bib}

\end{document}